\documentclass[reprint, preprintnumbers, superscriptaddress, prd, aps, amsmath,amssymb, twocolumn, a4paper, floatfix, noeprint]{revtex4-2}

\usepackage{dcolumn}
\usepackage{bm}
\usepackage[utf8]{inputenc}
\usepackage{graphicx,psfrag}
\usepackage{mathrsfs}
\usepackage{amsmath,amsfonts,amssymb}
\usepackage{multirow}
\usepackage{diagbox}
\usepackage{comment}
\usepackage{xcolor}
\usepackage{enumerate}
\usepackage{booktabs}
\usepackage[normalem]{ulem}
\usepackage{mathtools}
\usepackage{siunitx}
\usepackage{textgreek}
\usepackage[section]{placeins}
\usepackage{orcidlink}

\usepackage{color}
\definecolor{cyan}{rgb}{0,0.9,0.9}
\definecolor{orange}{rgb}{0.9,0.5,0}
\definecolor{magenta}{rgb}{1,0,1}
\definecolor{purple}{rgb}{0.8,0.4,0.8}
\definecolor{gray}{rgb}{0.8242,0.8242,0.8242}
\definecolor{green}{rgb}{0.,0.8,0.}

\usepackage{hyperref}
\hypersetup{
    colorlinks = true,
    linkcolor = {blue},
    citecolor = {blue},
    urlcolor = {blue},
    linkbordercolor = {white},
    breaklinks=true,
    }

\newcommand{\msun}{\text{M}_{\odot}}
\newcommand{\nsat}{\textit{n}_{\text{sat}}}
\newcommand{\fchi}{f_{\chi}}
\newcommand{\mchi}{m_{\chi}}
\newcommand{\mchimin}{m_{\chi, \text{min}}}
\newcommand{\fchimax}{f_{\chi, \text{max}}}

\newcommand{\mtov}{M_{\text{TOV}}}
\newcommand{\fmiq}{\text{fm}^{-3}}

\newcommand{\MeV}{\, \text{MeV}}
\newcommand{\sat}{\mathrm{sat}}
\newcommand{\sym}{\mathrm{sym}}

\begin{document}

\preprint{LA-UR-24-28505}

\title{Impact of dark matter on tidal signatures in neutron star mergers with the Einstein Telescope}  

\author{Hauke \surname{Koehn}\,\orcidlink{0009-0001-5350-7468}}
\email{hauke.koehn@uni-potsdam.de}
\affiliation{Institut f\"ur Physik und Astronomie, Universit\"at Potsdam, Haus 28, Karl-Liebknecht-Straße 24/25, 14476, Potsdam, Germany}

\author{Edoardo Giangrandi\,\orcidlink{0000-0001-9545-466X}}
\affiliation{CFisUC, Department of Physics, University of Coimbra, Rua Larga P-3004-516, Coimbra, Portugal}
\affiliation{Institut f\"ur Physik und Astronomie, Universit\"at Potsdam, Haus 28, Karl-Liebknecht-Str. 24/25, 14476, Potsdam, Germany}

\author{Nina Kunert\,\orcidlink{0000-0002-1275-530X}}
\affiliation{Institut f\"ur Physik und Astronomie, Universit\"at Potsdam, Haus 28, Karl-Liebknecht-Str. 24/25, 14476, Potsdam, Germany}

\author{Rahul \surname{Somasundaram}\,\orcidlink{0000-0003-0427-3893}}
\affiliation{Department of Physics, Syracuse University, Syracuse, New York 13244, USA}
\affiliation{Theoretical Division, Los Alamos National Laboratory, Los Alamos, New Mexico 87544 USA}

\author{Violetta Sagun\,\orcidlink{0000-0001-5854-1617}}
\affiliation{CFisUC, Department of Physics, University of Coimbra, Rua Larga P-3004-516, Coimbra, Portugal}

\author{Tim \surname{Dietrich}\,\orcidlink{0000-0003-2374-307X}}
\affiliation{Institut f\"ur Physik und Astronomie, Universit\"at Potsdam, Haus 28, Karl-Liebknecht-Str. 24/25, 14476, Potsdam, Germany}
\affiliation{Max Planck Institute for Gravitational Physics (Albert Einstein Institute), Am M{\"u}hlenberg 1, Potsdam 14476, Germany}

\begin{abstract}
If dark matter (DM) accumulates inside neutron stars (NS), it changes their internal structure and causes a shift of the tidal deformability from the value predicted by the dense-matter equation of state (EOS). In principle, this shift could be observable in the gravitational-wave (GW) signal of binary neutron star (BNS) mergers.
We investigate the effect of fermionic, noninteracting DM when observing a large number of GW events from DM-admixed BNSs with the precision of the proposed Einstein telescope (ET).
Specifically, we study the impact on the recovery of the baryonic EOS and whether DM properties can be constrained.
For this purpose, we create event catalogs of BNS mock events with DM fraction up to 1\%, from which we reconstruct the posterior uncertainties with the Fisher matrix approach.
Using this data, we perform joint Bayesian inference on the baryonic EOS, DM particle mass, and DM particle fraction in each event.
Our results reveal that when falsely ignoring DM effects, the EOS posterior is biased toward softer EOSs, though the offset is rather small. 
Further, we find that within our assumptions of our DM model and population, ET will likely not be able to test the presence of DM in BNSs, even when combining many events and adding Cosmic Explorer (CE) to the next-generation detector network.
Likewise, the potential constraints on the DM particle mass will remain weak because of degeneracies with the fraction and EOS.

\end{abstract}

\maketitle

\date{\today}
\section{Introduction}
\label{sec:introduction}
The detection of gravitational waves (GW) from compact binary mergers involving neutron stars (NSs) has provided the opportunity to constrain the equation of state (EOS) for dense, strongly interacting matter~\citep{LIGOScientific:2018cki, Radice:2017lry, Most:2018hfd, Raithel:2019uzi, Margalit:2017dij, Rezzolla:2017aly}, also in combination with additional recent data from nuclear and astrophysics~\citep{Raaijmakers:2021uju, Koehn:2024set, Huth:2021bsp, Capano:2019eae, Biswas:2021yge, Annala:2021gom}.
Due to their finite size, NSs experience tidal distortion if placed in an inhomogeneous gravitational field. During the inspiral of a binary neutron star (BNS) or black-hole-neutron-star binary (BHNS), this distortion causes a phase shift on the emitted GW signal. 
When this phase shift is measured, Bayesian inference allows for the recovery of the tidal deformability parameter $\Lambda$ which in turn restricts the underlying EOS for NSs. 

The usual assumption in the aforementioned studies is to consider compact stars in a vacuum. 
Yet, it has been suggested that NSs might be embedded in a dark matter (DM) halo or rapidly accrete DM clumps while passing through an overdense region in a subhalo~\citep{Bramante:2021dyx}. 
The latter scenario is supported by many cosmological models predicting formation of primordial DM gravitationally collapsed objects residing in subhalos~\citep{Erickcek:2011us,Buckley:2017ttd}. 
In fact, because of their extreme gravity, NSs may accumulate a significant amount of DM over their lifetime~\citep{Bose:2022ola, Robles:2022llu, Bertone:2007ae, deLavallaz:2010wp}, especially when located close to the respective galactic center~\cite{DelPopolo:2020hel}.
Furthermore, it is possible that upon the creation of the NS, DM is inherited from the progenitor star~\citep{Spolyar:2007qv, Meyer:2020vzy, Chan:2023atg}. 
In particular, BNS systems could have a relatively high amount of DM, since they are old systems that went through several stages of stellar evolution and the joint gravitational pull of the binary traps more DM particles in comparison to an isolated NS~\citep{Bell:2020jou}. 

DM in the NS interior would alter the interior structure and thereby impact properties such as radius, mass, and tidal deformability~\citep{Deliyergiyev:2019vti, Bertone:2007ae, Karkevandi:2021ygv, Das:2020ecp, Leung:2022wcf, Giangrandi:2022wht, Jockel:2023rrm}, for a recent review see Ref.~\citep{Bramante:2023djs}.
This makes NS potential probes for DM~\citep{Emma:2022xjs} but it also complicates the analysis in regards to the EOS for strongly interacting matter, because more unknown parameters with possible degeneracies need to be considered~\cite{Sagun:2022ezx,Giangrandi:2022wht}.
In Ref.~\cite{Rutherford:2022xeb}, the authors examined how future mass-radius measurements of NS with next-generation X-ray telescopes can be used to detect bosonic asymmetric dark matter, finding that with present uncertainties about the baryonic EOS, constraining intrinsic DM properties will not be feasible.
In Ref.~\cite{Sagun:2023rzp} it was shown that the reported low mass and radius for HESS J1731-347 can be explained by asymmetric, noninteracting DM and several values for the particle masses and fractions for DM would be possible.

In the present article, we investigate how future tidal deformability measurements of DM-admixed BNS with next-generation GW observatories impact the inference of the underlying baryonic EOS and whether they can be used to study intrinsic properties of DM.
Next-generation observatories are necessary for this purpose, since the precision of current GW measurements is not sufficient to identify the small offset in the tidal deformability induced by DM. 
With two detected BNS mergers~\citep{LIGOScientific:2017vwq, LIGOScientific:2020aai} and three BHNS candidates \citep{LIGOScientific:2021qlt, LIGOScientific:2024elc}, detector sensitivity limits have only allowed for meaningful constraints on $\Lambda$ in the case of GW170817~\citep{Raaijmakers:2021uju, Koehn:2024set}. 
Depending on the vicinity and rate of occurring events, tighter EOS constraints may be achieved during the advanced LIGO-Virgo observational runs O4 and O5~\citep{Finstad:2022oni, Bandopadhyay:2024zrr}, though the next major leap in precision will arise from next-generation detectors, such as the proposed Einstein telescope (ET)~\citep{Hild:2010id, Punturo:2010zz, Sathyaprakash:2012jk, Branchesi:2023mws} or Cosmic Explorer (CE)~\citep{LIGOScientific:2016wof, Reitze:2019iox, Evans:2021gyd}. 
Several studies for next-generation observations show that these facilities will be able to infer the EOS with high accuracy~\citep{Pacilio:2021jmq, Finstad:2022oni, Walker:2024loo, Bandopadhyay:2024zrr, Puecher:2023twf, Iacovelli:2023nbv}. 
In particular, the effective tidal deformability
\begin{align}
    \tilde{\Lambda} = \frac{16}{13} \frac{(1+12 q) \Lambda_1 + (12+q) q^4 \Lambda_2}{(1+q)^5}
\end{align}
in a binary with mass ratio $q$ and component deformabilities $\Lambda_{1,2}$ will be measurable with an uncertainty of $\sim \mathcal{O}(10)$ at the 90\% credibility limit~\citep{Puecher:2023twf, Puecher:2022oiz, Branchesi:2023mws, Iacovelli:2023nbv}. For comparison, the 90\%-credible interval from GW170817 is set at $\tilde{\Lambda} = 356^{+479}_{-217}$~\citep{LIGOScientific:2018mvr}. 
Hence, the question arises how to disentangle the contributions from DM and the unknown EOS, when tidal deformabilities can be determined with high precision in next-generation detectors.

For this purpose, we create BNS mock events where the tidal deformabilities are determined by an injected EOS together with DM effects taken into account.
The posteriors on $\tilde{\Lambda}$ that would be recovered by the next-generation GW observatories are approximated through the Fisher matrix approach~\cite{Vallisneri:2007ev}.
This mock posterior data is then used to perform a subsequent Bayesian analysis to infer the baryonic EOS jointly with the parameters of the DM model.
We focus mainly on BNS measurements with ET, while we provide conclusions for a joint detector network of ET and CE in the Appendix.

We consider asymmetric, noninteracting, fermionic DM that does not annihilate, and, therefore, steadily accumulates inside the compact stars.
Asymmetric DM is consistent with the ΛCDM model predicting cold (nonrelativistic) collisionless DM particles and reproduces the existing observational data~\citep{Planck:2018vyg}. 
The amount of accrued DM in BNSs depends on the surrounding medium, particularly the DM particle density, and could differ for each system~\citep{Kouvaris:2010jy}. 
This aspect poses a complication to the analysis of the observational data, but also offers the potential to firmly detect DM when significant variation in the tidal signatures is observable after comparing many events.
The present study addresses this question by considering BNS merger events with different DM fractions up to 1\%, reflecting deviations of DM density in each galaxy and at different distances from the galactic center~\citep{DelPopolo:2020hel}.

The present article is organized as follows: In Sec.~\ref{sec:EOS_construction}, we describe our baryonic EOS set and how we add the effects of DM for the determination of the tidal deformability. 
Thereafter in Sec.~\ref{sec:EOS_inference}, we describe how we create the mock event data and lay out the details of the subsequent Bayesian statistical analysis. 
In Sec.~\ref{sec:results}, we present our findings for the impact of DM on the inference of the EOS and assess whether ET will be able to distinguish between populations of BNSs with and without DM.
Finally, we summarize and discuss our findings in Sec.~\ref{sec:conclusion}.

\section{Constructing NS with DM}
\label{sec:EOS_construction}
\subsection{EOS construction}
\begin{table}[t]
\centering
\tabcolsep=0.3cm
\def\arraystretch{1.5}
\caption{The distributions from which the empirical parameters are drawn to generate the EOS candidates. The parameters E$_\sat$ and $\nsat$ are fixed at \qty{-16}{MeV} and \qty{0.16}{\fmiq}, respectively. We denote uniform distributions by $\mathcal{U}$.}
\begin{tabular}{>{\centering\arraybackslash} p {3.5cm} >{\centering\arraybackslash} p {3.5 cm}}
 \toprule
 \toprule
 Parameter & Distribution\\
 \midrule
 $K_{\sat}$ [\unit{MeV}] & $\mathcal{U}(210, 260)$\\
 $Q_{\sat}$ [\unit{MeV}]& $\mathcal{U}(-1000,1000)$\\
 $Z_{\sat}$ [\unit{MeV}]& $\mathcal{U}(-1000,1000)$\\
 \midrule 
 $E_{\sym}$ [\unit{MeV}]& $\mathcal{U}(28,35)$ \\
 $L_{\sym}$ [\unit{MeV}]& $\mathcal{U}(30,100)$ \\
 $K_{\sym}$ [\unit{MeV}]& $\mathcal{U}(-200,200)$ \\
 $Q_{\sym}$ [\unit{MeV}]& $\mathcal{U}(-1000,1000)$ \\
 $Z_{\sym}$ [\unit{MeV}]& $\mathcal{U}(-1000, 1000)$\\
\bottomrule
\end{tabular}
\label{tab:NEP_range}
\end{table}
We construct our set of nucleonic EOSs using the metamodel of Refs.~\citep{Margueron:2017eqc,Margueron:2017lup}. 
The construction is similar to that of Ref.~\citep{Koehn:2024set} but with the difference that, in this work, we use the metamodel at all densities, i.e., we do not allow for the presence of non-nucleonic degrees of freedom in NSs. 
The EOS predicted by the metamodel is determined by the nuclear empirical parameters which are given by the expansion of the energy per particle in symmetric nuclear matter $\mathcal{E}(n)$, and the symmetry energy $\mathcal{S}(n)$, around the nuclear saturation density, $n_{\rm sat}$,  
\begin{align}
\begin{split}
    &\mathcal{E}(n)=E_\text{sat}+\frac{1}{2}K_\text{sat}x^2+\frac{1}{6}Q_\text{sat}x^3\\ & \quad \quad \quad +\frac{1}{24}Z_\text{sat}x^4 + \dots\,,
\end{split}\\
\begin{split}
    &\mathcal{S}(n)=E_\text{sym}+L_\text{sym}x+\frac{1}{2}K_\text{sym}x^2+\frac{1}{6}Q_\text{sym}x^3 \\ & \quad \quad \quad +\frac{1}{24}Z_\text{sym}x^4 + \dots\,,
\end{split}
\end{align}
where $x=\frac{n-n_\text{sat}}{3n_\text{sat}}$. 
In this work, we fix $E_\text{sat} = -16\,\MeV$ and $n_\text{sat} = 0.16\,\fmiq$. 
The higher-order isoscalar parameters are varied uniformly in the ranges specified in Table~\ref{tab:NEP_range}. Similarly, Table~\ref{tab:NEP_range} also contains the ranges for the isovector parameters.  
By drawing the nuclear empirical parameters from the specified ranges, we create a large set of EOS candidates from the metamodel.
Furthermore, we require that our EOS candidates satisfy the following constraints:
\begin{enumerate}[(i)]
    \item A maximum mass of $\mtov > 1.97\,\msun$ to account for the radio observations of heavy pulsars~\citep{Demorest:2010bx, NANOGrav:2019jur, Fonseca:2021wxt}.
    \item Positive symmetry energy $\mathcal{S}(n)>0$ at all NS densities, see Ref.~\citep{Margueron:2017lup}.
    \item Causality in the speed of sound, $c_s \leq c$  for $n < 1.2 \,n_\textrm{TOV}$, where $n_\textrm{TOV}$ is the central density of a NS with the maximum mass $\mtov$.  
\end{enumerate}
Metamodel draws that do not satisfy the above criteria are discarded. In this manner, we generate a set of 5000 nucleonic EOSs that constitute our EOS prior. 

\subsection{Adding DM to the NS properties}
\begin{table}[t]
\renewcommand{\arraystretch}{1.1}
\caption{Grid values for the DM particle mass $\mchi$ and DM fraction $\fchi$ for which NS configurations were determined. In total this amounts to $12\times12$ combinations. The values are distributed log-uniformly.}
\label{tab:DM_parameters}
\begin{minipage}[t]{0.2\textwidth}
\centering
\begin{tabular}{>{\centering\arraybackslash}p{2cm}}
\toprule
\toprule
$\mchi$ in \MeV \\
\midrule
170 \\ 
221 \\
286 \\
372 \\
483 \\
627 \\
814 \\
1056 \\
1371 \\
1780 \\
2311 \\
3000 \\
\bottomrule
\end{tabular}
\end{minipage}
{\huge $\times$}
\begin{minipage}[t]{0.2 \textwidth}
\centering
\begin{tabular}{>{\centering\arraybackslash}p{2 cm}}
\toprule
\toprule
$\fchi$ in \% \\
\midrule
0.01 \\ 
0.015 \\
0.023 \\
0.035 \\
0.053 \\
0.081 \\
0.123 \\
0.187 \\
0.285 \\
0.433 \\
0.658 \\
1 \\
\bottomrule
\end{tabular}
\end{minipage}
\end{table}
We model the DM component as a relativistic Fermi gas of noninteracting spin-one-half particles minimally coupled to Standard Model particles. This model has been extensively studied in the literature, for instance in Refs.~\citep{Nelson:2018xtr,Ivanytskyi:2019wxd}. Hence, the DM pressure is given by
\begin{align}
\begin{split}
    p(\mu_{\chi}) = \frac{\mchi^4 c^5}{24 \pi^2 \hbar^3} \bigg(& x \sqrt{x^2-1} (2 x^2 - 5) \\
     & + 3 \log\left(x+\sqrt{x^2-1}\right) \bigg)\,,
\end{split}
\end{align}
where $x$ is the ratio of the chemical potential $\mu_{\chi}$ to the DM particle mass $\mchi$.By adopting the noninteracting Fermi gas model, we keep the number of parameters needed to describe DM to just two: the particle mass and fraction, while preserving the essential physics. In this way, the fermionic pressure is sufficient to balance the gravitational pull of DM, allowing for stable configurations and preventing DM-admixed NSs from collapsing into black holes, aligning with the observation of old compact stars.

The relation between NS mass, radius, and tidal deformability is determined by solving the two-fluid Tolman–Oppenheimer–Volkoff (TOV) and Love equations, discussed in detail in~\citep{Ivanytskyi:2019wxd}. 
Note, that the star's tidal deformability is obtained considering the effective speed of sound derived in Ref.~\citep{Giangrandi:2022wht}.
Unlike the one-fluid case, where the mass-radius curve is solely determined by the EOS, here the possible NS configurations will also depend on the DM particle mass $\mchi$ and the fraction $\fchi$ of (gravitational) DM mass over the total (gravitational) mass of the NS.
For each of our 5000 baryonic EOS candidates, we calculate the stable NS configurations on a grid of $12\times12$ combinations for the DM particle masses and fractions.
The possible grid values are distributed log-uniformly between \qtyrange{170}{3000}{MeV} and $0.01$--$1\%$ and listed in Table~\ref{tab:DM_parameters}.
The two-fluid TOV equations are solved up to the point where the NS mass starts to decrease with higher central density to ensure stability of the NS against radial perturbations~\cite{Routaray:2022utr}.

The choice of the DM particle mass range is motivated by previous studies~\citep{Ivanytskyi:2019wxd,Ruter:2023uzc} where it was shown that light fermionic DM with $\mchi \lesssim \qty{170}{MeV}$ forms an extended halo around the NS. 
Similar configurations could be obtained for most bosonic DM models, the range of parameters across different models is essentially equivalent~\cite{Karkevandi:2021ygv}.
Such extended DM halos could overlap at the late inspiral phase of the merger which makes existing waveform models inapplicable (for details see~\citep{Ruter:2023uzc}).
We thus focus on DM core configurations, considering only small changes in tidal deformability and assuming that the waveform description remains otherwise unaltered.
As typical DM fractions for NS remain unknown, the choice for the range of $\fchi$ remains arbitrary, though for instance Ref.~\citep{Ivanytskyi:2019wxd} argue that considering only the Bondi accretion of DM onto an NS close to the Galactic center could yield $\fchi = 0.01\%$. 
Higher fractions could occur through the other mechanisms mentioned in Sec.~\ref{sec:introduction} and we consider 1\% as a realistic upper limit.
We note that other works also have investigated NS configurations with higher fractions~\cite{Rutherford:2022xeb, Miao:2022rqj, Giangrandi:2022wht}.

\section{EOS inference through BNS mergers}
\label{sec:EOS_inference}
\begin{figure}[t!]
    \centering
    \includegraphics[width = \linewidth]{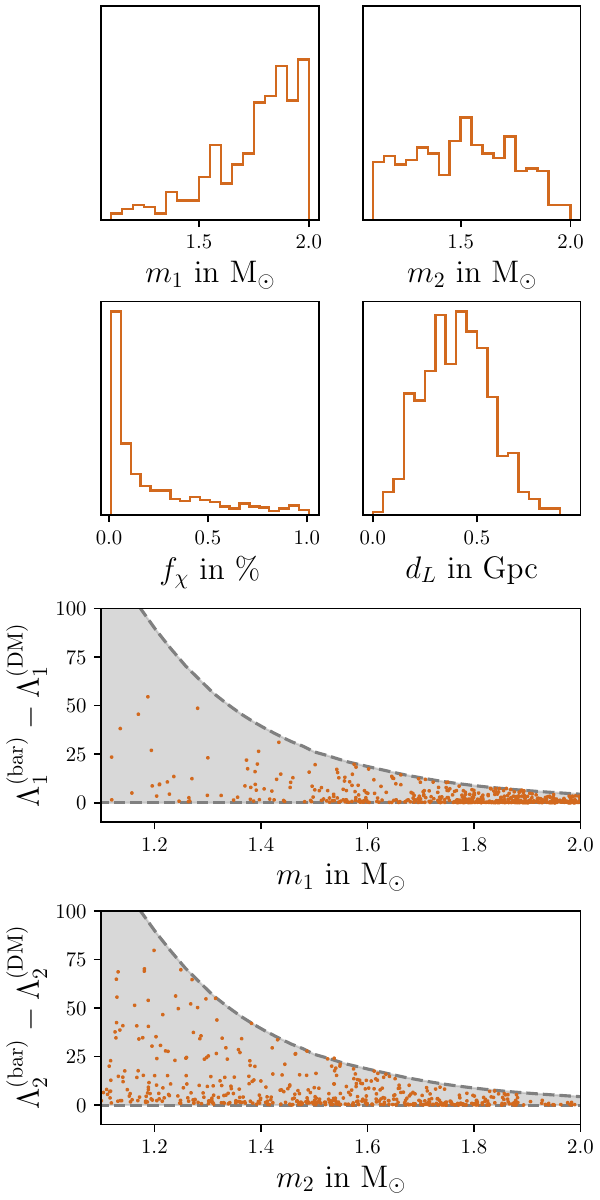}
    \caption{Overview of the 500 mock BNS events. The upper panels show the distribution of the component masses $m_1$, $m_2$, and luminosity distance $d_L$, these values remain the same across all event catalogs.
    The bottom panels show the difference between $\Lambda^{\text{(bar)}}$, i.e., the tidal deformability expected from the baryonic EOS, and the actual tidal deformability when accounting for the presence of DM. The upper dashed gray line shows the limit on the deviation of the tidal deformability from the baryonic value, i.e., from the $M$-$\Lambda$ relationship for a NS with 1\% DM.
    For this plot, we chose the event catalog with the stiff baryonic EOS, $\mchi=\qty{483}{MeV}$ and the fraction $\fchi$ was distributed log-uniformly.}
    \label{fig:mock_events}
\end{figure}

For our analysis, we create sixteen distinct catalogs of mock ET events, each encompassing 500 BNS events with a signal-to-noise ratio (SNR) larger than 100.
The catalogs are identical in terms of the underlying population model, i.e., the NS masses and distances stay the same, but differ in terms of the injected baryonic EOS, DM particle mass, or DM fraction population. 
Specifically, for each catalog we pick one baryonic EOS, one value for $\mchi$ and decide whether the DM fraction $\fchi$ is sampled log-uniformly or uniformly between $10^{-4}$ and $10^{-2}$. 
In total, we inject two different EOSs with four possible values $\mchi$ (\qty{170}{\MeV}, \qty{483}{\MeV}, \qty{1056}{\MeV}, or \qty{3000}{\MeV}), which then with the choice for the $\fchi$ distribution amounts to $2\times4\times2$ combinations that represent our sixteen different catalogs. 
We show the distribution of the event parameters for one of our catalogs in Fig.~\ref{fig:mock_events}.
For the purpose of background checking, we also create two event catalogs that do not contain any DM effects. In these catalogs, the tidal deformability is solely determined by the baryonic EOS. We refer to them as ``no DM" catalogs.

From these events, we create mock posteriors on the tidal deformability using \textsc{gwfast}~\citep{Iacovelli:2022mbg, Iacovelli:2022bbs} and use those to perform inference on our EOS candidate set.
The following two subsections first provide more details on how we draw the events for our catalogs and then about the statistical analysis of our Bayesian inference of the EOS.

\subsection{Creating mock events}
\begin{table}[t!]
    \centering
    \caption{Distributions from which the parameters for the individual events are drawn. Our event catalogs are created by sampling from these distributions and discarding any event with an SNR lower than 100. The event catalogs only differ in the choice for the EOS, $\mchi$ and the distribution for $\fchi$. The draws for the masses, spins, and observational parameters remain the same.}
    \label{tab:population_distribution}
    \begin{tabular}{clrlr}
    \toprule
    \toprule
       &   parameter  &   \multicolumn{2}{c}{symbol} & \multicolumn{1}{c}{distribution} \\
      \midrule
      \multirow{5}{8.5pt}{\rotatebox{90}{intrinsic}} & Component mass $[\rm M_\odot] $ &&$m_1, m_2$ &  Eq.~\eqref{eq:BNS_masses}\\
        & Spin magnitude && $a_{1}, a_{2}$ & $\mathcal{U}(0,0.05)$\\
        & DM fraction && $\fchi$ & $\text{Log or }\mathcal{U}(10^{-4}, 10^{-2})$\\

        & Tidal deformability &&$\Lambda_1, \Lambda_2$ & from EOS, $\mchi, \fchi$ \vspace{0.5 cm}\\
       \multirow{6}{8.5pt}{\rotatebox{90}{observational}}& Luminosity distance [Mpc] &&$d_{L}$ & $\mathcal{U}_{\text{com. vol.}}(1,1000)$ \\
       & Sky position [rad] && $\varphi, \theta$ & $\mathcal{U}(0, 2\pi)$, Cos$(0,\pi)$ \\
       & Trigger time [GPS] && $t_c$ & $\mathcal{U}(\qty{1}{yr})$\\
       & Phase [rad] &&$\phi_c$ & $\mathcal{U}(0,2\pi)$\\
       & Inclination [rad] && $\iota$& Cos$(0,\pi)$\\
       & Polarization [rad] &&$\psi$ & $\mathcal{U}(0,2\pi)$\\
       \bottomrule
    \end{tabular}
\end{table}
To create a mock GW detection for ET, we need to specify a parameter vector $\vec{\theta}$ for each event. 
This parameter vector encompasses the masses and tidal deformabilities of the NSs, their spins and other observational parameters. 
We list the distributions from which we draw each quantity in Table~\ref{tab:population_distribution}. 
The mass function for merging BNSs is largely unconstrained, therefore, we use the following distribution 
\begin{align}
\begin{split}
    &P(m_1, m_2) \propto \\
    &\begin{cases} (m_2/m_1)^2  \qquad \text{if} \quad 1.1\,\msun \leq m_2 < m_1\leq 2\,\msun \\
    0 \qquad \qquad \qquad \text{else}
    \end{cases} \,,
\end{split}
\label{eq:BNS_masses}
\end{align}
based on the analysis of GW170817 and GW190425 in Ref.~\citep{Landry:2021hvl}. 

The tidal deformabilities are determined through the combination of baryonic EOS, $\mchi$ and $\fchi$. 
We inject one of two baryonic EOSs that were arbitrarily chosen from our candidate set. 
We refer to one as the "stiffer" EOS. 
For a canonical \qty{1.4}{\msun} NS without any DM this EOS predicts a tidal deformability of $\Lambda_{1.4}^{(\text{bar})} = 534$.
The superscript here indicates that it is the expectation from the baryonic EOS alone and no DM effects are taken into account.
The other EOS is the "softer" one with $\Lambda_{1.4}^{(\text{bar})} = 293$. 
The EOS and $\mchi$ are fixed in each catalog, the DM fraction $\fchi$ varies between individual events. 
We assume that both NSs in a binary have the same fraction $\fchi$ that is either drawn from a log-uniform distribution between 0.01\% and 1\%, or from a uniform distribution in the same range. 

To draw conclusions about the EOS from a BNS signal $d_{\text{GW}}$, one has to consider the posterior $P(\tilde{\Lambda}, \mathcal{M}, \eta|d_{\text{GW}})$ for the (source frame) chirp mass $\mathcal{M}$, symmetric mass ratio $\eta$, and effective tidal deformability $\tilde{\Lambda}$. This posterior is usually inferred from a full Bayesian sampling with a likelihood given by 
\begin{align}
\begin{split}
   \ln \mathcal{L}(\vec{\theta}|d_{\text{GW}}) =& -2 \int_{f_{\rm min}}^{f_{\rm max}} \frac{\lvert d_{\text{GW}}(f) - h(f,\vec{\theta})\rvert^2 }{S(f)} \ df\,.
   \label{eq:GW_likelihood}
\end{split}
\end{align}
Here, $h(f, \vec{\theta})$ is the waveform model prediction for the parameter point $\vec{\theta}$, and $S(f)$ denotes the detector power spectral density. 
However, especially for long high-SNR signals, sampling over the entire parameter space turns out to be rather expensive.
Instead, for the type of loud signals we are interested in, we assume that $P(\tilde{\Lambda}, \mathcal{M}, \eta|d_{\text{GW}})$ is well approximated by a Gaussian centered around the injected (true) value. 
The Fisher matrix 
\begin{align}
    F_{jk} = \mathbb{E}\left(\frac{\partial \ln \mathcal{L}(\vec{\theta} |d_{\text{GW}}) }{\partial \theta_j}   \frac{\partial \ln \mathcal{L}(\vec{\theta} |d_{\text{GW}}) }{\partial \theta_k}\right)\,,
\label{eq:FI}
\end{align} 
is then used to construct the uncertainty of this posterior.
Specifically, the covariance matrix for $P(\tilde{\Lambda}, \mathcal{M}, \eta|d_{\text{GW}})$ is the inverse $F_{jk}$. 
To determine Eq.~\eqref{eq:FI} and its inverse, we use the \textsc{gwfast} package \citep{Iacovelli:2022mbg, Iacovelli:2022bbs}. The waveform model for the determination of the likelihood was set to $\mathtt{IMRPhenomD\_NRTidalv2}$ \citep{Husa:2015iqa, Khan:2015jqa, Dietrich:2019kaq}. 
For the ET detector, we assume the triangle configuration located in Sardinia \citep{Puecher:2023twf, Branchesi:2023mws}, the power spectral density was taken from Ref.~\citep{ET_D_PSD}.

When assessing the detection probabilities of DM, it is necessary to compare against the case when no DM is present in the events. 
For this reason, we also create two event catalogs in a similar manner as described here, but the tidal deformabilities are solely determined by the baryonic EOS. 
For each of these two ``no DM" catalogs we use either the stiff or soft EOS from above.

\subsection{Analyzing the events}
The probability to observe a posterior $P(\tilde{\Lambda}, \mathcal{M}, \eta|d_{\text{GW}})$ from an event $d_{\text{GW}}$ when taken an EOS, a value $\fchi$ for the DM fraction, and a value $\mchi$ for the DM particle mass as given, is proportional to
\begin{align}
\begin{split}
    \mathcal{L}(\text{EOS}, \mchi, \fchi|d_{\text{GW}}) &= \\
    &\int d\mathcal{M}\ d\eta\ P(\tilde{\Lambda}(\mathcal{M}, \eta), \mathcal{M}, \eta|d_{\text{GW}})\,.
    \label{eq:DM_likelihood}
\end{split}
\end{align}
Here, the relationship $\tilde{\Lambda}(\mathcal{M}, \eta)$ is determined through the EOS, $\fchi$, and $\mchi$. 

We are interested in the posterior probability for an EOS given a mock observation. From the likelihood in Eq.~\eqref{eq:DM_likelihood} it can be derived as 
\begin{align}
\begin{split}
    &P(\text{EOS}|d_{\text{GW}}) = \int d\mchi\ d\fchi\ P(\text{EOS}, \mchi, \fchi|d_{\text{GW}}) \\
     &\propto \int d\mchi\ d\fchi\ \mathcal{L}(\text{EOS}, \mchi, \fchi|d_{\text{GW}})\ \pi(\text{EOS}, \mchi, \fchi)\,.
     \label{eq:EOS_posterior}
\end{split}
\end{align}
For the prior $\pi(\text{EOS}, \mchi, \fchi)$, we assume all the baryonic EOSs to be of equal prior likelihood, whereas the prior in $\mchi$ is taken to be log-uniformly distributed between \qtyrange{170}{3000}{MeV}. 
The prior for $\fchi$ is chosen to match the distribution used to generate the events, i.e., either log-uniform or uniform between \qtyrange{0.01}{1}{\%}. 
Therefore, we may approximate Eq.~\eqref{eq:EOS_posterior} by taking the log-uniform samples $\{m_{\chi, j}\}$ and $\{f_{\chi,k}\}$ from Table~\ref{tab:DM_parameters} and sum over them
\begin{align}
    P(\text{EOS}|d_{\text{GW}}) \approx  \frac{1}{N_{\mchi} N_{\fchi}} \sum_{j, k} \mathcal{L}(\text{EOS}, m_{\chi, j}, f_{\chi,k}|d_{\text{GW}}).
\end{align}
The number of samples for $\mchi$ and $\fchi$ from Table~\ref{tab:DM_parameters} is $N_{\mchi} = N_{\fchi} = 12$.
In those cases where we want a uniform prior on $\fchi$, the summation over the fraction samples is performed with corresponding weights.
It is worth pointing out that when combining multiple events $d^{(1)}_{\text{GW}}$, $d^{(2)}_{\text{GW}}$, one has to marginalize over the fraction for each event individually, i.e.,
\begin{align}
&\mathcal{L}(\text{EOS}, m_{\chi,j}|d^{(1)}_{\text{GW}}) \approx \frac{1}{N_{\fchi}} \sum_k \mathcal{L}(\text{EOS}, m_{\chi,j}, f^{(1)}_{\chi, k}|d_{\text{GW}}^{(1)})\,,\\
\begin{split}
    &\mathcal{L}(\text{EOS}, m_{\chi,j}|d^{(1)}_{\text{GW}}, d^{(2)}_{\text{GW}}) = \\
    &\qquad \qquad \mathcal{L}(\text{EOS}, m_{\chi,j}|d^{(1)}_{\text{GW}})\ \mathcal{L}(\text{EOS}, m_{\chi,j}|d^{(2)}_{\text{GW}}) \,. \label{eq:combined_likelihood}
\end{split}
\end{align}
The posterior on the EOS is simply obtained by marginalizing the likelihood with a log-uniform prior on the DM particle mass:
\begin{align}
\begin{split}
P(\text{EOS}|d^{(1)}_{\text{GW}}, d^{(2)}_{\text{GW}}) \propto 
\frac{1}{N_{\mchi}} \sum_{j} \mathcal{L}(\text{EOS}, m_{\chi,j}|d^{(1)}_{\text{GW}}, d^{(2)}_{\text{GW}})\,.
\label{eq:combined_EOS_posterior}
\end{split}
\end{align}
To obtain a posterior on the DM particle mass instead, we can also marginalize Eq.~\eqref{eq:combined_likelihood} with a uniform prior on the baryonic EOS.

We establish two hypotheses, to explore the potential biases when analyzing the DM-admixed BNSs without accounting for the effects of DM:
\begin{itemize}
    \item $\mathcal{H}_0$: The null-hypothesis  states that for a given set of BNS events, the data is simply described by one baryonic EOS. 
    \item $\mathcal{H}_1$: The alternative hypothesis claims that the BNS events are best described by DM-admixed BNS and the $\tilde{\Lambda}(\mathcal{M}, \eta)$ relationship experiences some variation from event to event. 
\end{itemize}

The hypothesis $\mathcal{H}_1$ corresponds to using the likelihood in Eq.~\eqref{eq:DM_likelihood}, whereas for $\mathcal{H}_0$ the likelihoods need to be determined with
\begin{align}
    \mathcal{L}(\text{EOS}|d_{\text{GW}}) = \int d\mathcal{M}\ d\eta\ P(\tilde{\Lambda}^{\text{(bar)}}(\mathcal{M}, \eta), \mathcal{M}, \eta|d_{\text{GW}})\,,
\end{align}
i.e., the relationship for the tidal deformability is simply given by the baryonic EOS without any DM effects.

\section{Results}
\label{sec:results}
For each event in our 16 catalogs, we determine the likelihood in Eq.~\eqref{eq:DM_likelihood} across our EOS candidates and the parameter grid for $\mchi$ and $\fchi$ from Table~\ref{tab:DM_parameters}.
From these likelihood values we obtain the posterior distribution on the baryonic EOS as described in Eq.~\eqref{eq:combined_EOS_posterior}. Similarly, we also obtain a posterior on the DM particle mass $\mchi$ and further determine the Bayes factor for $\mathcal{H}_1$ vs.\ $\mathcal{H}_0$.
In this section, we summarize the findings of our Bayesian analysis.
\subsection{Recovering the underlying baryonic EOS}
\label{subsec:tidal_recovery}
\begin{figure*}
    \centering
    \includegraphics[width=\linewidth]{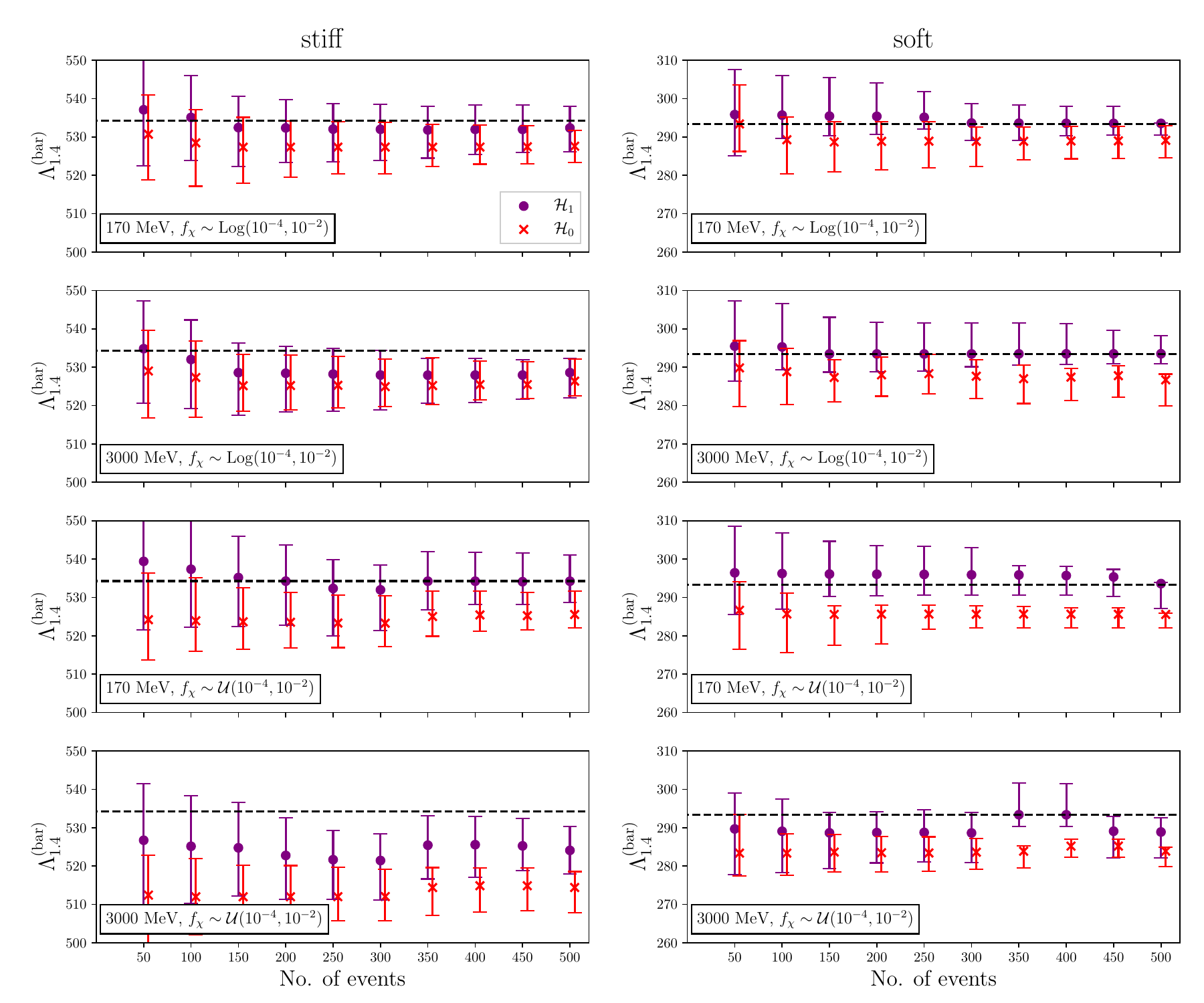}
    \caption{Evolution of the posterior estimate for the baryonic $\Lambda_{1.4}^{(\text{bar})}$ with rising number of observed  BNS events. The purple bands show the 95\% credibility interval when the events are analyzed according to $\mathcal{H}_1$, i.e., when DM effects are included, whereas the red band indicates the credibility interval if DM effects are not accounted for, corresponding to $\mathcal{H}_0$. The circle and cross indicate the posterior medians, respectively. The panels correspond to distinct event catalogs, where the injected DM particle mass and the population for the DM fraction are described in the bottom left label. The panels of the left side correspond to those catalogs, where the injected baryonic EOS is stiffer, the panels on the right to the softer EOS. The true value for $\Lambda_{1.4}^{(\text{bar})}$ from the injected baryonic EOS is shown as black dashed line.}
    \label{fig:event_overview}
\end{figure*}
For the DM model and DM parameter ranges we consider in the present article, the tidal deformabilities of BNS are systematically lowered. 
Hence, over many events a bias toward softer baryonic EOSs could arise if one were to analyze a set of DM-admixed BNS events simply through a baryonic set of EOS candidates. 
Our analyses confirm that small biases in the EOS recovery appear, if the events in a catalog with DM are analyzed according to $\mathcal{H}_0$, i.e., ignoring DM effects. 
Figure~\ref{fig:event_overview} compares how the posterior estimate on the canonical baryonic tidal deformability $\Lambda_{1.4}^{(\text{bar})}$ evolves over the course of multiple BNS events, when the DM effects are included in the inference vs.\ when they are ignored.
Note that $\Lambda_{1.4}^{(\text{bar})}$ here refers to the tidal deformability without any DM effects, as we simply use this quantity as an indicator to describe the recovery of the injected baryonic EOS.

The strength of the bias depends on whether the population for the DM fraction is log-uniform or uniform. 
In case of the latter, more BNS events will have a higher $\fchi$ and already after around \qtyrange{50}{100}{} events, a discrepancy between the $\mathcal{H}_0$-analyses and the injected true EOS is noticeable. At the same time, this bias is generally small, as the posterior estimates deviates from the true $\Lambda_{1.4}^{(\text{bar})}$ in the range of \qtyrange{10}{20}{}. 
When the DM fraction in the BNS events is log-uniformly distributed, the discrepancy between the injected value and the analysis without DM is even smaller and the 95\% credible region excludes the true EOS only after $\gtrsim 150$ events. 
The analyses for the event catalogs with injected $\mchi = \qty{483}{MeV}$, and $\mchi=\qty{1056}{MeV}$ are not shown in Fig.~\ref{fig:event_overview} but show the same trend.

However, we note that even the $\mathcal{H}_1$-inference is not always successful at recovering the true EOS. 
This happens in those cases where the injected particle mass is high, as seen in Fig.~\ref{fig:event_overview} for those catalogs where $\mchi = \qty{3000}{MeV}$. 
The reason is that the core configurations get more compact with higher DM particle mass, hence the reduction of the tidal deformability from the pure baryonic EOS is larger than for lower $\mchi$. 
However, this reduction in $\Lambda$ can also be mimicked by EOSs that are softer than the actual injected EOS when they are combined with a low DM particle mass. 
Because we marginalize over all DM particle masses, and lower masses are preferred in the log-uniform prior, the final result is biased toward these slightly softer EOSs. 
The bias disappears when using a prior that is more favorable for higher $\mchi$. 
This indicates the sensitivity of the $\mathcal{H}_1$-inferences to the prior choice for the DM parameters. 

Conversely to the bias we receive when $\mathcal{H}_1$ is true but we instead assume $\mathcal{H}_0$ in our analysis, we can also look at the bias that arises when we consider the events without any DM and analyze them according to $\mathcal{H}_1$.
For this purpose we reran the analysis for our two ``no DM" catalogs.
Analyzing these events with $\mathcal{H}_0$, we found no bias at all in the recovery of the true baryonic EOS, as expected. 
When performing the analysis with $\mathcal{H}_1$, the posterior on $\Lambda_{1.4}^{(\text{bar})}$ is biased toward stiffer EOSs than the true one, though the offset is only \qtyrange{5}{15}{} and the true injected value leaves the 95\% credible interval only after 400 events.

We point out that the uncertainties with which we determine $\Lambda_{1.4}^{(\text{bar})}$ in the catalogs with the softer baryonic EOS (right panels in Fig.~\ref{fig:event_overview}), are likely underestimated. 
This is because the softer EOS lies in a region of our EOS prior that is not very densely sampled, so after $\gtrsim 300$ events only few EOS with nonzero likelihood remain. 
Hence, the posterior on $\Lambda_{1.4}^{(\text{bar})}$ is undersampled and the 95\% credible intervals become too narrow. 

\subsection{Detectability of DM}

\begin{figure*}
    \centering
    \includegraphics[width=\linewidth]{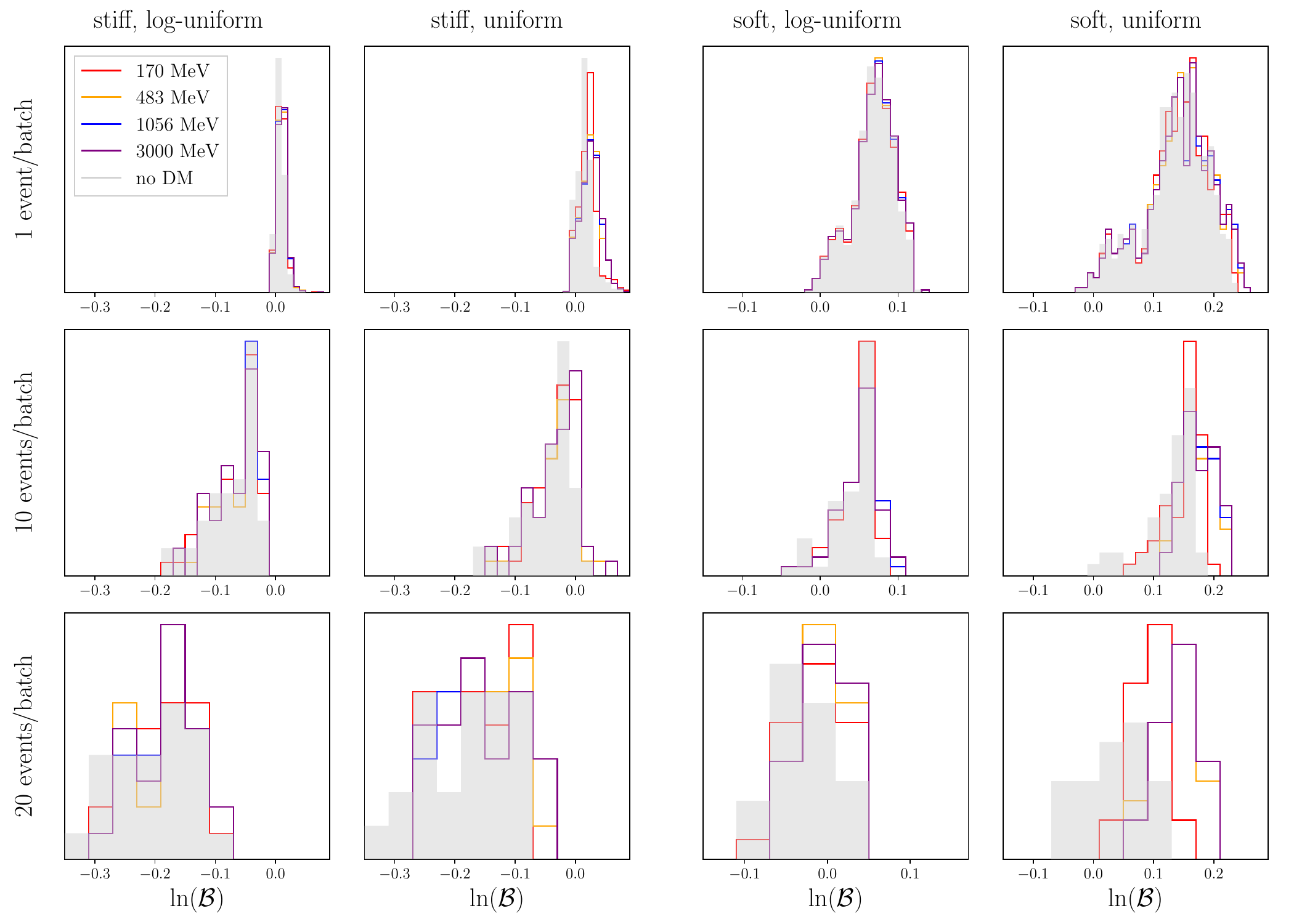}
    \caption{Distribution of $\ln(\mathcal{B})$ for the sixteen event catalogs with DM and the two event catalogs without DM.
    The top panels show how the Bayes factor is distributed when analyzing each of the 500 events individually, the middle panels show the distribution for batches of 10 events, the bottom panels for batches of 20 events.
    The first panel column refers to the event catalogs where the stiff EOS is used and $\fchi$ is distributed log-uniformly. 
    The different colors indicate the different injected $\mchi$-values, the gray patch refers to the evidence distribution when no DM is present in the events. 
    Likewise, the second panel column uses the stiff EOS and a uniform $\fchi$-distribution, the third the soft EOS and a log-uniform $\fchi$-distribution, and the fourth the soft EOS and a uniform $\fchi$-distribution.}
    \label{fig:evidence_distribution}
  
\end{figure*}

\begin{table}
    \centering
    \caption{Log-Bayes factors $\ln(\mathcal{B})$ for $\mathcal{H}_1$ against $\mathcal{H}_0$ after 500 events. The first column specifies which baryonic EOS was used in the respective event catalog, the second one whether $\fchi$ was distributed log-uniform or uniform. The log-Bayes factors are listed in the columns for each injected $\mchi$ and for the no-DM catalogs.}
    \begin{tabular}{p {1 cm} p { 1.7 cm} >{\centering\arraybackslash}p {1 cm} >{\centering\arraybackslash}p { 1  cm} >{\centering\arraybackslash}p {1 cm} >{\centering\arraybackslash}p { 1 cm} >{\centering\arraybackslash}p { 1 cm}}
    \toprule
    \toprule
    EOS  & distribution for $\fchi$ &  \multicolumn{4}{c}{$\mchi$ in MeV} &  \\
      &  &  $170$ & $483$ & $1056$ & $3000$ & no DM\\
    \midrule
stiff & $\mathrm{Log}$ & $-4.54$ & $-4.32$ & $-4.23$ & $-4.23$ & $-4.09$\\ 
stiff & $\mathcal{U}$ & $-3.78$ & $-2.41$ & $-2.23$ & $-2.16$ & $-3.68$\\ 
& & & & & \\
soft & $\mathrm{Log}$ & $-1.98$ & $-2.56$ & $-2.62$ & $-2.71$ & $-2.53$\\ 
soft & $\mathcal{U}$ & $-2.92$ & $-2.93$ & $-2.70$ & $-2.65$ & $-1.79$\\ 
    \bottomrule
    \end{tabular}
    \label{tab:evidences}
\end{table}
To assess whether ET would allow the detection of DM from our BNS population, one may look at the Bayes factor $\mathcal{B}$ for $\mathcal{H}_1$ against $\mathcal{H}_0$.
It is given as the evidence ratio of $\mathcal{H}_1$ over $\mathcal{H}_0$.
We approximate the evidences by averaging the likelihood over the prior samples for $\mchi$, $\fchi$, and the EOS.
The Bayes factor will change, depending on how many events are combined in the likelihood.

In the top panels of Fig.~\ref{fig:evidence_distribution}, we determine $\ln(\mathcal{B})$ for each event individually and plot the distribution. 
The individual log-Bayes-factors range mostly from -0.02 to 0.25.
Thus, a single event is not very indicative whether DM was present in the NS or not.
This is expected, as a single event only provides one data point for $\tilde{\Lambda}$ that likely can be well matched by some baryonic EOS.

However, even when combining multiple events together, the detectability of DM does not increase in our setup.
In the middle and bottom panels of Fig.~\ref{fig:evidence_distribution}, we randomly group 10, respectively 20 events together, determine the Bayes factor for these batches and show the distribution.
Generally, with growing number of combined events, the Bayes factor decreases and becomes mostly negative.
The reason is that we average over the likelihood values with our prior to determine the evidence, and when multiple likelihoods from different events are combined, the average over the larger parameter space will generally decrease quicker than over the smaller one, when both models have the same ability to describe the data.
The preference for the simpler model is a general property of the Bayes factor often attributed to Occam's razor.
We note that the distribution of $\ln(\mathcal{B})$ is rather independent of the value for $\mchi$.
When comparing the Bayes factors in a forward-background approach, we see that ET is not capable to distinguish between $\mathcal{H}_1$ and $\mathcal{H}_0$.
The two ``no DM" catalogs, i.e., the catalogs for which $\mathcal{H}_0$ is actually true, serve as background model and the distribution of their log-Bayes-factors are shown as gray patches in the panels of Fig.~\ref{fig:evidence_distribution}. 
They cover a similar range as the evidence ratios for the catalogs where DM is present, indicating that one cannot decide $\mathcal{H}_1$ over $\mathcal{H}_0$ based on $\mathcal{B}$.
This is definitely true when $\fchi$ is distributed log-uniformly, whereas $\ln(\mathcal{B})$ sometimes rises above the maximum from the background catalog when the fraction instead follows a uniform distribution.
However, $\ln(\mathcal{B})$ stays close to 0 and for neither DM catalog it raises above the 5--$\sigma$ level estimated from the background model with kernel-density estimation.
For the case of the soft EOS, there is an additional caveat: 
As discussed in Sec.~\ref{subsec:tidal_recovery}, the soft EOS lies in a slightly undersampled region of the EOS prior. 
Therefore, the corresponding analyses assume stronger \textit{a priori} knowledge about the EOS and are thus more capable of distinguishing between the tidal deformabilities from the baryonic and DM cases. 
The slightly stronger separation of $\ln(\mathcal{B})$ in case of the catalogs with the soft EOS indicates that potent constraints about the baryonic EOS, e.g., from nuclear physics or other data points, may enable detection of DM through tidal information from BNS, though it is unclear to which extent the baryonic EOS would need to be constrained.
Yet, when inferring the EOS and DM properties simultaneously as we do in the present work, combining up to 20 BNS events will not be sufficient to observe the presence of DM from tidal signatures of BNS.

In Table~\ref{tab:evidences} we show the values for $\ln(\mathcal{B})$ when combining all 500 events together.
For all catalogs, the log-Bayes factor is negative, indicating that even with 500 events, $\mathcal{H}_0$ is equally capable of describing the data as $\mathcal{H}_1$. 
The expected number of BNS detections with SNR $>100$ for ET is hard to assess due to uncertainties in the BNS population and the eventual detector design, but estimates range from \qtyrange{50}{150} per year~\citep{Borhanian:2022czq, Evans:2021gyd, Branchesi:2023mws}. 
Hence, we expect that ET will not be able to identify DM from tidal signatures of BNS within its first years of operation.
In the Appendix we confirm that this assessment does not change, when the same set of events is analyzed in a detector network with ET and CE.

\subsection{Inference of the DM particle mass}
\begin{figure}
    \centering
    \includegraphics[width= \linewidth]{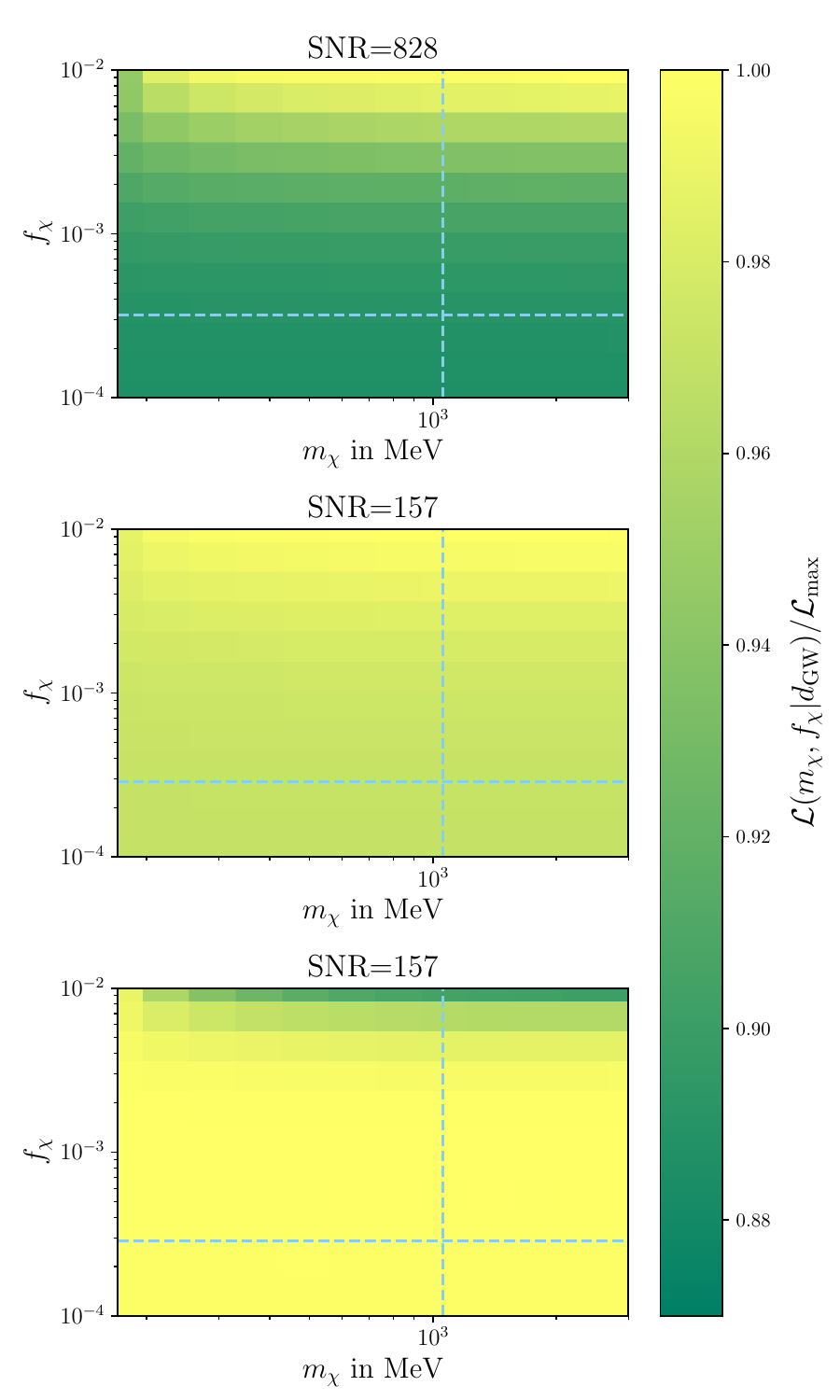}
    \caption{Likelihood across the $\mchi$-$\fchi$ plane after individual events. The top panel shows the likelihood after one event with high SNR where $\mchi = \qty{1056}{MeV}$ and $\fchi =0.032\%$ (both values shown as dashed blue lines), when marginalizing over all EOS candidates. Similarly, the middle panel shows an event with a modest SNR where $\mchi = \qty{1056}{MeV}$ and $\fchi =0.029\%$. The bottom panel shows the same event as in the middle panel, but here the likelihood is just taken from the true EOS to showcase the degeneracy between low $\mchi$ and high $\mchi$ with small fractions.}
    \label{fig:mchi_fchi_distribution}
\end{figure}
\begin{figure}
    \centering
    \includegraphics[width= \linewidth]{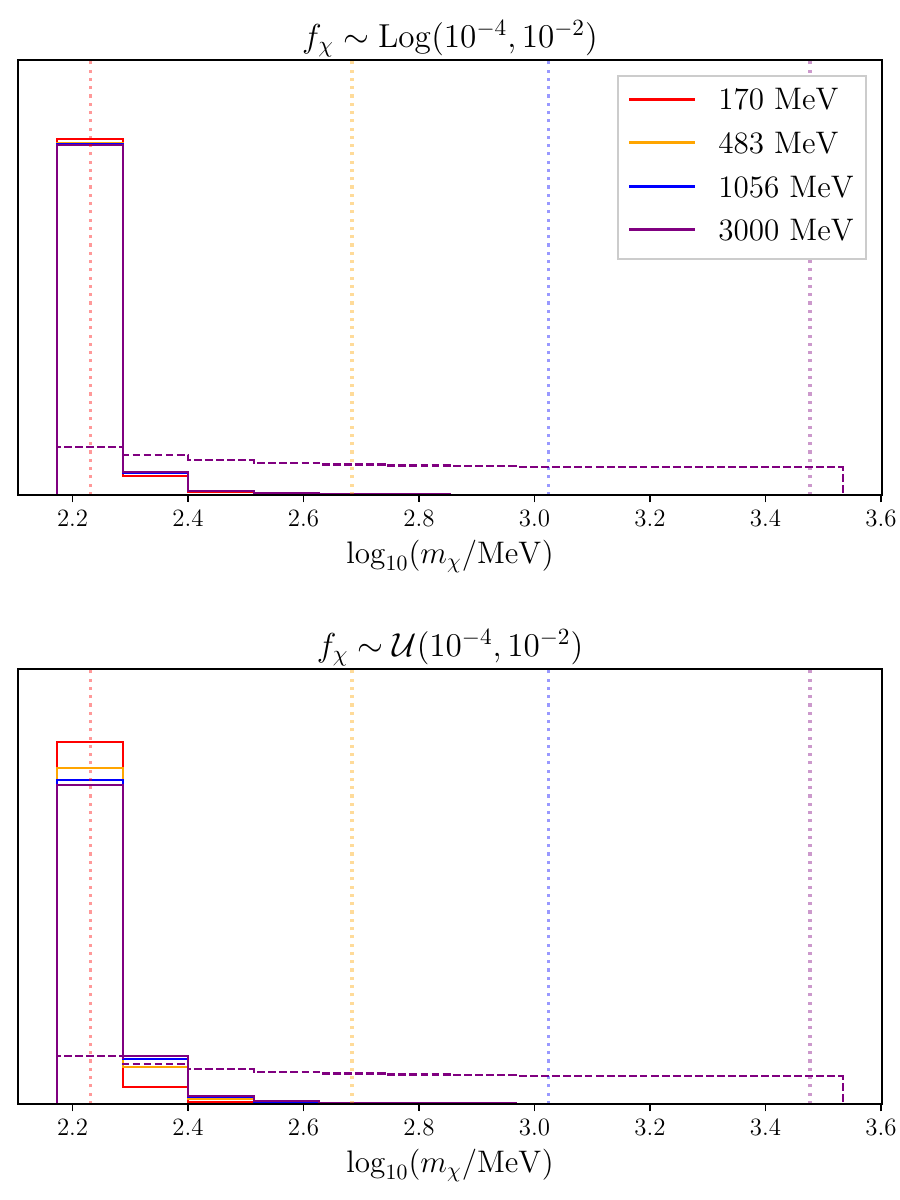}
    \caption{Posterior distributions on $\mchi$ after 500 events. The colors indicate which DM particle mass was injected in the respective event catalog. The four different injection values are also drawn as vertical dotted lines. We only show event catalogs with the stiff EOS, the top panels considers those where $\fchi$ was log-uniformly distributed, the bottom panel the ones where $\fchi$ is uniform. The purple dashed lines show the posterior after 50 events with an injected DM particle mass of \qty{3000}{MeV} to illustrate that the posterior skew to low masses is a cumulative effect from the degeneracy mentioned in the text.}
    \label{fig:mchi_posterior}
\end{figure}
Since the DM particle mass impacts the change in tidal deformability of the NS, one could in principle infer the value of $\mchi$ from the GW data. 
In practice however, we find that this is difficult to achieve when inferring the EOS and DM properties simultaneously.
An individual event is only one data point about the tidal deformability, hence getting a useful estimate on $\mchi$ and $\fchi$ is infeasible.
In Fig.~\ref{fig:mchi_fchi_distribution}, we show the distribution on $\mchi$ and $\fchi$ after one event, in one instance after the event with the highest SNR and in the other for an event with average SNR. When marginalizing over all prior EOS candidates, i.e.,
\begin{align}
    \mathcal{L}(\mchi, \fchi | d_{\text{GW}}) = \sum_{\text{EOS}} \mathcal{L}(\mchi, \fchi, \text{EOS} | d_{\text{GW}})\,.
\end{align}
the likelihood is relatively constant across the $\mchi$-$\fchi$ space, even for the high SNR event. 
Higher masses and fractions seem to have slightly higher likelihoods, though the variation is only on the order of a few percent.
This offset of the maximum likelihood from the true values arises because DM reduces the tidal deformabilities in the signal by an interval $\approx \Delta\Lambda(\mchi, \fchi)$. 
The likelihood for a pair of $\mchi$ and $\fchi$ thus depends on how many baryonic EOSs are available within the range of the tidal deformability as measured plus $\Delta\Lambda(\mchi, \fchi)$. 
Our EOS prior set is not exactly uniform in tidal deformability and in this case there are just slightly more baryonic EOSs available to match the data when the values for $\mchi$ and $\fchi$ are high compared to when they are low, hence the offset arises.

To get an estimate of the DM particle mass, multiple events are needed to infer the true baryonic EOS simultaneously with the deviation in the observation away from it.
In the bottom panel of Fig.~\ref{fig:mchi_fchi_distribution} we show the likelihood for $\mchi$ and $\fchi$ after the event with modest SNR, where we only consider the contribution from the true injected EOS. 
On the one hand, it demonstrates that, when the true underlying EOS is known, the maximum likelihood aligns well with the true values of $\mchi$ and $\fchi$. 
However, the variation in likelihood is still small, and even for high SNR events we find that it is only on the order of a couple $\sim10\%$. 
Moreover, the bottom panel of Fig.~\ref{fig:mchi_fchi_distribution} also exhibits a degeneracy between $\mchi$ and $\fchi$.
An event with high DM particle mass and low fraction may be equally well described by a low DM particle mass. 
When the DM particle mass is large, but the fraction small, the shift in tidal deformability from the baryonic EOS is only slight. 
The likelihood of the large particle mass will only be high when the fraction is small. 
But the likelihood for the smaller DM particle mass will be high regardless the value of $\fchi$. 
By restricting the DM fraction to below 1\%, the resulting DM profile for light DM particles does not exceed the baryonic radius, meaning that halo configurations are not possible. 
Instead, the formed DM core causes only a small shift in tidal deformability, which remains within the measurement uncertainty.

This is an inherent problem when trying to infer the DM particle mass from multiple events.
Our event catalogs contain many events with small DM fraction.
Since we marginalize over the fraction for each event individually, the events with low fraction will have a small preference for lower $\mchi$, as they are able to explain the data across all fractions.
The preference is only very small, but can accumulate over many events, which after $\gtrsim 100$ events skews the combined posterior heavily toward lower masses.
This effect can be seen in Fig.~\ref{fig:mchi_posterior} for the posterior on $\mchi$ after 500 events.
Events with high fraction can correct the estimate again toward larger masses, however, even when the population of $\fchi$ is uniform, there are not enough events with high fraction to overcome the opposite trend toward small $\mchi$.
We point out that this problem remains even if the baryonic EOS was perfectly known, i.e., if one does not marginalize the likelihood Eq.~\eqref{eq:combined_likelihood} over the EOSs, but instead just considers the contribution from the true injected one. 

This means that tidal signatures of BNS events alone are not sufficient to measure $\mchi$.
However, we briefly note a method here how to deduce a lower limit $\mchimin$ on the DM particle mass.
Namely, when we assume that the maximum DM-fraction a BNS can have is $\fchimax$,
and we consider a particular baryonic EOS together with DM particle masses that yield core configurations,  we can determine the minimum effective tidal deformability $\tilde{\Lambda}_{\text{min}}$ a BNS can have. 
This will be a function of the chirp mass $\mathcal{M}$, symmetric mass ratio $\eta$ and additionally depends on the EOS, as well as the DM parameter bounds $\mchimin$ and $\fchimax$, i.e.,
\begin{align}
    \tilde{\Lambda}_{\text{min}}(\mathcal{M}, \eta) = \tilde{\Lambda}(\mathcal{M}, \eta, \text{EOS}, \mchimin, \fchimax)\,.
    \label{eq:lambdat_min}
\end{align}
Now, when many events have been observed, one can select the single event $d_{\text{GW}}^{(\star)}$ where the observed $\tilde{\Lambda}$ deviates the most from the expectation of the baryonic EOS. 
It is reasonable to assume that it had the maximum DM fraction $\fchimax$, so we can then derive the probability that $\mchi$ is smaller than the boundary $\mchimin$ as 
\begin{align}
\begin{split}
    &P(\mchi \leq  \mchimin | d^{(\star)}_{\text{GW}}, \fchimax, \text{EOS}) =\\
    & \int d\mathcal{M} \int d\eta \int_{\tilde{\Lambda}_{\text{min}}(\mathcal{M}, \eta)}^\infty d\tilde{\Lambda}\ P(\mathcal{M}, \eta, \tilde{\Lambda} | d^{(\star)}_{\text{GW}})\,.
    \label{eq:mchimin_likelihood}
\end{split}
\end{align}
The lower bound in the last integral is determined by Eq.~\eqref{eq:lambdat_min}.
Typically, the higher one sets $\fchimax$, the lower $\mchimin$ can be. 
This is because a larger deviation from the baryonic EOS can also be explained by small DM particle masses when the DM fraction is high enough. 

Note that this approach requires assumptions about the baryonic EOS and the maximum DM fraction in NS. However, one could for instance marginalize Eq.~\eqref{eq:mchimin_likelihood} over the EOS posterior from the 500 events and likewise marginalize with a prior on $\fchimax$.
An additional practical issue concerns the determination which event $d_{\text{GW}}^{(\star)}$ has the highest deviation from the baryonic EOS.
For our purposes here, we simply look at the posterior mean $\tilde{\Lambda}$ from the data and compare it to the value determined from the baryonic EOS alone with the posterior mean for $\mathcal{M}$ and $\eta$. 
We point out that in practice, determining $d_{\text{GW}}^{(\star)}$ is potentially more difficult, since for real data the posterior mean does not always need to be well centered and other methods to identify $d_{\text{GW}}^{(\star)}$ could be more suitable.
Our results however show that the constraints on $\mchimin$ remain weak anyway. 
Even when the true value for $\mchi$ is \qty{3000}{MeV}, we still find $P(\mchi \leq \qty{170}{MeV}|d^{(\star)}_{\text{GW}}, \fchimax = 1\%)= 34\%$.
This indicates again that even the events that show the strongest signs for DM only restrict the particle mass weakly.

\section{Discussion and Conclusions}
\label{sec:conclusion}
In the present article, we studied how ET measurements of the tidal deformability of DM-admixed NS can be used to recover the baryonic EOS and infer properties of DM.
To this end, we created a catalog of 500 high-SNR BNS events and used the Fisher matrix approach to obtain estimates of the posterior uncertainties. 
In different instances of the catalog the injected baryonic EOS, injected DM particle mass, and the chosen distribution for the DM fraction were varied.

Using these event catalogs as our mock dataset, we performed joint Bayesian inference on the EOS, $\mchi$, and $\fchi$. 
We find that DM in BNS can bias the inference of the baryonic EOS when its presence is not accounted for, though the effect is very small.
The offset in the tidal deformability $\Lambda_{1.4}^{(\text{bar})}$ between the true baryonic EOS and the EOS posterior depends on the injected value for $\mchi$ as well as the population for $\fchi$, but is generally $\lesssim 20$.
Similarly, the posterior on the canonical NS radius $R_{1.4}$ likewise is only off up to $\lesssim \qty{0.1}{km}$.
Note, that the effect could be stronger for heavier DM particles and higher fractions. 

Even when DM is present in the data, we generally observe that the hypothesis for the presence of DM is disfavored by the Bayes factor when combining many events. 
This is because the observational signatures of DM are not strong and hence over many events the easier hypothesis requiring only the baryonic EOS is preferred.
Comparison to a background analysis where the events actually lack DM reveals that the Bayes factor is not able to distinguish between the two cases. 
This is even true for the more favorable setups where $\fchi$ is distributed uniformly instead of log-uniformly, though we also find indication that if the true EOS was perfectly known, ET might be able to deliver some evidence about the presence of DM.

Likewise, inferring the DM particle mass seems infeasible because of an inherent degeneracy between $\fchi$ and $\mchi$.
When the particle mass is large and the fraction low, the data may be equally well described by a small particle mass and higher DM fraction within the measurement uncertainty.
Alternatively, we attempted to construct a lower limit on $\mchi$ based on the assumed maximum DM fraction a BNS can have.
In our setup though, this does not deliver any pertinent result either, since the statistical uncertainties on the event that deviates the most from the purely baryonic expectation are still too large.

The conclusions drawn in the present work depend on the choice of our DM model as well as on the assumptions about the ranges for the DM particle mass and DM fraction.
Stronger observational signatures could be achieved with a different DM model, higher fractions, and different particle masses. 
Based on our results though, firm detection of DM would require a much larger variation of the NS tidal deformability than we consider in our event catalogs.
Alternatively, if the baryonic EOS is well-constrained from other independent sources, DM detection might be possible.
We leave the investigation into the necessary variability of tidal deformability or the required accuracy on the baryonic EOS for confident DM detection to future work.

While our analysis can be seen as cautious for the range of $\fchi$ and $\mchi$, we point out that our statistical analysis benefits from some advantages a real-life application would not have.
Namely, the NSs in our binary events always share the same fraction and we imposed the correct prior for $\fchi$ based on the population distribution that we injected in the event catalog.
Further, the Fisher matrix is generally considered a lower bound on the statistical uncertainties, and in a full Bayesian parameter estimation the uncertainty on $\tilde{\Lambda}$ can be significantly larger~\citep{Iacovelli:2023nbv}.
It has been shown that the Fisher matrix approach can also \textit{overestimate} the uncertainty on certain parameters~\cite{Rodriguez:2013mla}, though this mainly applies when parameters display mutual degeneracies or when the Fisher matrix implies support within a nonphysical range, e.g., negative values for the luminosity distance~\cite{Dupletsa:2024gfl}.
However, in our event catalogs, the Fisher matrix uncertainties on $\tilde{\Lambda}$ cover a range between \qtyrange{4}{80}, which agrees well with posterior estimates from full Bayesian parameter estimations for ET~\citep{Puecher:2023twf, Iacovelli:2023nbv}.
Hence, we do not expect that our Fisher matrix approach overestimates the accuracy with which ET would recover the tidal deformability, especially when additionally considering systematic uncertainties in the waveform models~\cite{Kunert:2021hgm, Read:2023hkv, Samajdar:2018dcx, Owen:2023mid, Gamba:2020wgg}.
Nevertheless, full Bayesian parameter estimation could be implemented in future to test if biases in other parameters arise when DM effects in BNSs are ignored.

Reduction of the tidal deformability in NS is not the only way DM could be observable through future GW telescopes. 
With the precision of next-generation detectors, several alternative possibilities for DM detection have been suggested, e.g., an additional feature in the BNS postmerger signal~\citep{Ellis:2017jgp,Suarez-Fontanella:2024epb}, continuous GW signals around superradiant black holes~\cite{Baumann:2022pkl}, observing sub-2-$\msun$ black holes from DM-collapsed NS~\cite{Singh:2022wvw, Bhattacharya:2023stq, Dasgupta:2020mqg}, or direct interaction of DM particles with the detector test masses~\cite{Pierce:2018xmy, Tsuchida:2019hhc, Chen:2021apc}.
However, based on our analysis we conclude that within our assumptions about the DM  particle mass and DM fractions in NS, confident detection of DM cores through tidal measurements of BNSs by ET is unlikely.

\begin{acknowledgments}

We thank Francesco Iacovelli and Michele Mancarella for helpful assistance with the \textsc{gwfast} package.

H.~K., N.~K., and T.~D. acknowledge funding from the Daimler and Benz Foundation for the project “NUMANJI”. This work was co-funded by the European Union (ERC, SMArt, 101076369). Views and opinions expressed are those of the authors only and do not necessarily reflect those of the European Union or the European Research Council. Neither the European Union nor the granting authority can be held responsible for them.

The work of V.~S. and E.~G. was supported by national funds from FCT – Fundação para a Ciência e a Tecnologia within the projects UIDP/\-04564/\-2020 and UIDB/\-04564/\-2020, respectively, with DOI identifiers 10.54499/UIDP/04564/2020 and 10.54499/UIDB/04564/2020. E.~G. also acknowledges the support from Project No. PRT/BD/152267/2021.

R.S. acknowledges support from the Nuclear Physics from Multi-Messenger Mergers (NP3M) Focused Research Hub which is funded by the National Science Foundation under Grant Number 21-16686, and from the Laboratory Directed Research and Development program of Los Alamos National Laboratory under project number 20220541ECR.

The authors gratefully acknowledge the Gauss Centre for Supercomputing e.V. (www.gauss-centre.eu) for funding this project by providing computing time on the GCS Supercomputer SuperMUC-NG at Leibniz Supercomputing Centre (www.lrz.de).
\end{acknowledgments}

\appendix*
\section{INCLUDING COSMIC EXPLORER IN THE DETECTION NETWORK}
\begin{figure}
    \centering
    \includegraphics[width=1\linewidth]{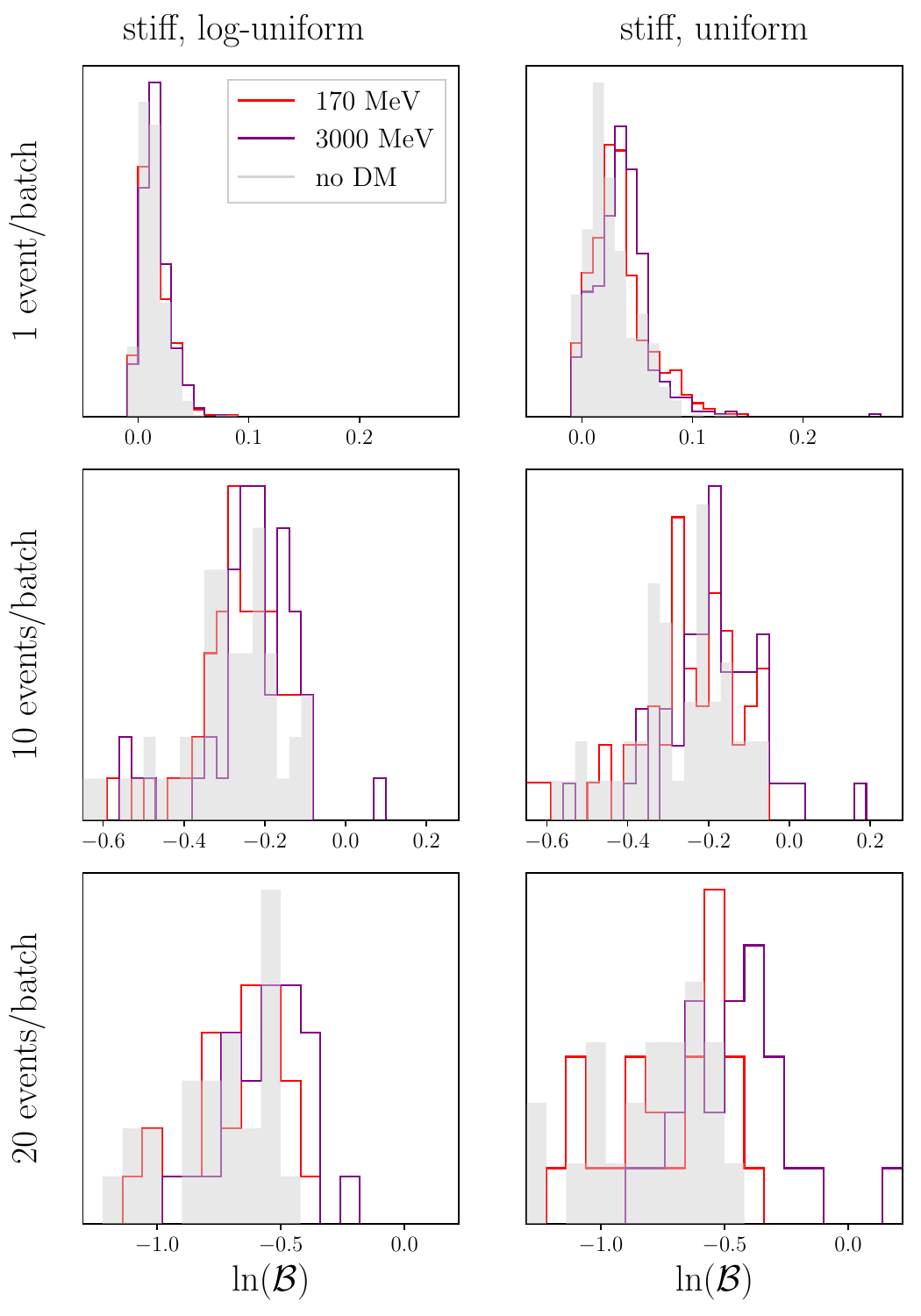}
    \caption{Distribution of $\ln(\mathcal{B})$ for the four event catalogs where we included CE into the detector network. Figure arrangement and color coding as in Fig.~\ref{fig:evidence_distribution}.}
    \label{fig:evidence_distribution_CE}
\end{figure}
\label{app:CE}
The inference of BNS parameters from GW signals will have notably lower uncertainties if ET is joined by the other proposed third-generation detector CE.
To see how our conclusions would change if CE was included in the detector network, we reran the analysis for some of our event catalogs. 
To obtain the reduced statistical uncertainties on $\mathcal{M}$, $\eta$, and $\tilde{\Lambda}$, one needs to simply add the Fisher matrices of the individual detectors
\begin{align}
    F_{jk} = F_{jk}^{(\text{ET})} + F_{jk}^{(\text{CE})}.
\end{align}
This is implemented directly in \textsc{gwfast}. 
We assume that the CE detector is located in Idaho, with the PSD taken from Ref.~\citep{CE_PSD}.
With this setup, we recreate the mock posteriors for the event catalogs where the stiff baryonic EOS was injected and $\mchi$ was set to \qty{170}{MeV} or \qty{3000}{MeV}. 
Hence, we consider ET and CE for $1\times2\times2$ DM catalogs plus one additional no-DM catalog.
The subsequent Bayesian inference of the EOS and DM parameters is conducted in the same manner described in Sec.~\ref{sec:EOS_inference}.

The recovery of the baryonic EOS shows similar behavior to Sec.~\ref{subsec:tidal_recovery}, though the statistical uncertainties on the recovered baryonic EOS are now even further reduced. 
In particular, undersampling effects in our baryonic EOS prior now become noticeable even for the stiff EOS after $\gtrsim300$ events.
The detectability of DM however is not improved by the reduced uncertainties on $\tilde{\Lambda}$, $\mathcal{M}$, and $\eta$. 
In Fig.~\ref{fig:evidence_distribution_CE}, we show the distribution of $\ln(\mathcal{B})$ for the four event catalogs we included CE in.
The forward-background approach still shows that the Bayes factor will not allow to distinguish whether DM was present within 20 events. 
Moreover, $\mathcal{B}$ decreases the more events are added, similar to the cases with ET only.
When we combine all events, excluding only the one with the highest SNR, we find $\ln(\mathcal{B})=-9.01$ ($-6.68$) for the catalog with \qty{170}{MeV} and where $\fchi$ distributed log-uniformly (uniformly). 
Likewise, the log-Bayes factors for the catalog with \qty{3000}{MeV} is $-6.77$ ($-5.27$) and $-9.48$ ($-7.82$) for the no-DM catalog.

We excluded here the one event with the highest SNR of $3158$ which originates from an event with two \qty{1.5}{\msun} NSs at \qty{41}{Mpc} distance.
Its DM fraction is 0.07\% and the Fisher matrix error on the tidal deformability lists at $\sigma_{\tilde{\Lambda}} = 1.4$. 
If this event was included, the log-Bayes factor for the event catalog with $\mchi=\qty{3000}{MeV}$ and $\fchi\sim\mathcal{U}(10^{-4}, 10^{-2})$ formally becomes 14.62 after all 500 events, correctly indicating positive support for $\mathcal{H}_1$.
We note that the event on its own does not yield the detection of DM, in fact the Bayes factor only starts showing substantial support for $\mathcal{H}_1$ when it is combined with $\gtrsim 40$ other events.

While it might be possible that particular "golden" events with very accurate determination of $\tilde{\Lambda}$ warrant the detection of DM when combined with a sufficient amount of other data about the baryonic EOS, we note a couple of caveats.
In the other event catalogs which include similar versions of this event,  the log-Bayes factor remains negative, indicating that the formally positive $\ln(\mathcal{B})$ originates from undersampling effects in our EOS prior, i.e., the likelihood is so constraining that only a few of our EOS candidate samples remain.
Hence, further research, preferably employing full Bayesian parameter estimation, seems needed to confirm if one "golden event" can indeed reverse the evidence.
Furthermore, in the real-life application, systematic errors in the waveform models could potentially prevent the extraction of $\tilde{\Lambda}$ at such accuracy~\cite{Kunert:2021hgm, Read:2023hkv, Samajdar:2018dcx, Owen:2023mid, Gamba:2020wgg}.

Overall, we therefore still conclude that even with a joint network of CE and ET, tidal BNS signatures would likely not enable the detection of the DM candidate adopted in the present work.

\vfill
\newpage
\bibliographystyle{prx.bst}
\bibliography{bibliography}
\end{document}